\documentclass[twocolumn,amsmath,amssymb,aps,preprintnumbers,showpacs,superscriptaddress]{revtex4}

\pdfoutput=1

\usepackage{ulem}
\usepackage{graphics}
\usepackage{dcolumn}
\usepackage{bm}
\usepackage{amsmath}
\usepackage{amssymb}
\usepackage{hypernat}
\usepackage[pdftex,hyperindex=true,final=true,colorlinks=true,pagebackref=false,bookmarks=true,bookmarksopen=true,bookmarksnumbered=true]{hyperref}
\usepackage{epsfig}

\newif\iffig
\newif\ifgraph

\figtrue                     %

\graphtrue                   %

\begin{document}


\title{Excess flux stability fingerprints in the $I_c(B)$-$T_c(B)$ phase boundary of superconducting thin films with quasiperiodic microtopology}

\author{D.~Bothner}\email{daniel.bothner@uni-tuebingen.de}
\affiliation{Physikalisches Institut -- Experimentalphysik II and Center for Collective Quantum Phenomena in LISA$^+$, Universit\"{a}t T\"{u}bingen, D-72076 T\"{u}bingen, Germany}
\author{R.~Seidl}
\affiliation{Physikalisches Institut -- Experimentalphysik II and Center for Collective Quantum Phenomena in LISA$^+$, Universit\"{a}t T\"{u}bingen, D-72076 T\"{u}bingen, Germany}
\author{V.~R.~Misko}
\affiliation{Departement Fysica, Universiteit Antwerpen, B-2020 Antwerpen, Belgium},
\author{R.~Kleiner}
\affiliation{Physikalisches Institut -- Experimentalphysik II and Center for Collective Quantum Phenomena in LISA$^+$, Universit\"{a}t T\"{u}bingen, D-72076 T\"{u}bingen, Germany}
\author{D.~Koelle}
\affiliation{Physikalisches Institut -- Experimentalphysik II and Center for Collective Quantum Phenomena in LISA$^+$, Universit\"{a}t T\"{u}bingen, D-72076 T\"{u}bingen, Germany}
\author{M.~Kemmler}
\affiliation{Physikalisches Institut -- Experimentalphysik II and Center for Collective Quantum Phenomena in LISA$^+$, Universit\"{a}t T\"{u}bingen, D-72076 T\"{u}bingen, Germany}

\date{\today}

\begin{abstract}

We experimentally investigate the magnetic field $B$ dependence of the critical current $I_c$ and the transition temperature $T_c$, \textit{i.e.} the $I_c(B)-T_c(B)$ phase boundary, of superconducting niobium thin films patterned with periodic and quasiperiodic antidot arrays on the submicron scale.
For this purpose we monitor current-voltage characteristics at different values of $B$ and $T$.
We investigate samples with antidots positioned at the vertices of two different tilings with quasiperiodic symmetry, namely the Shield Tiling and the Tuebingen Triangle Tiling.
For reference we investigate a sample with a triangular antidot lattice.
We find modulations of the phase boundary for both quasiperiodic tilings, which were predicted by numerical simulations but not observed in experiments yet.
The particularity of these commensurability effects is that they correspond to excess flux densities, which are slightly higher than the matching flux.
The observed matching effects can be explained by quasiperiodic caging of interstitial vortices and/or the formation of symmetry induced giant vortices.

\end{abstract}


\maketitle

\section{Introduction}

\label{sec:Introduction}
The topology of superconducting thin films has a strong impact on the shape of the superconductor-normalconductor phase boundary in an external magnetic field $B$.
Basically the reason for this connection is the fluxoid quantization condition, requiring that along each closed path inside the superconductor the gauge invariant phase is an integer multiple of $2\pi$.
As pointed out first by Little and Parks \cite{Little62} this quantization condition leads to an oscillation of the transition temperature $T_c$ of a superconducting cylinder in a magnetic field.
In a field parallel to the cylinder symmetry axis, the periodicity of the $T_c(B)$-oscillations is one flux quantum $\Phi_0=h/2e=2.07\times10^{-15}\,$Tm$^2$ per cross sectional area of the cylinder.
These Little-Parks oscillations were also found in two-dimensional (2D) superconducting wire networks with periodic or quasiperiodic structure \cite{Pannetier84, Behrooz86, Nori87}, where the flux quantization has not only to be fulfilled for each single loop of the network but also for all paths around multiple loops.
In these samples, commensurable states between the applied flux and the network are also observable at fractional (periodic networks) or even irrational (quasiperiodic networks) multiples of one flux quantum per single loop.
Another famous consequence of the fluxoid quantization is the formation of Abrikosov vortices, when a type-II superconducting thin film is exposed to a perpendicular magnetic field.
In an ideal type-II superconductor, these vortices with normal conducting cores are repulsively interacting and form a highly ordered state, the hexagonal Abrikosov vortex lattice \cite{Abrikosov57}.
As the formation of a normal conducting core costs condensation energy, any defect in a real superconductor, where the superconducting phase is weakened, can be viewed as an energy minimum for an Abrikosov vortex.
For the case of an ensemble of vortices interacting with an ensemble of defects, the repulsive interaction of the vortices plays an important role and the resulting spatial vortex configuration is determined by an interplay of the energy gain due to vortex pinning at defects and the energy cost due to the elastic vortex lattice deformation.
With modern lithography techniques it is possible to create lattices of artificial pinning sites, \textit{e.g.}, antidots \cite{Fiory78, Moshchalkov98}, magnetic dots \cite{Otani93, Martin97} or carbon nanotubes \cite{Haeffner09} with almost arbitrary topology, on the characteristic length scales of the superconductor (penetration depth $\lambda$ and coherence length $\xi$).
Such pinning potential landscapes can be varied in terms of size, geometry and amplitude during the patterning process.
In addition, the strength and range of the vortex-vortex as well as the vortex-defect interaction can be tuned during the measurements by varying the sample temperature $T$ and hence $\lambda$ and $\xi$.
Finally, it is possible to easily control the vortex density via the applied magnetic field and to create constant and alternating forces on the vortices by applying external currents.
Hence, Abrikosov vortices in a micropatterned superconducting film constitute a highly designable model system for interacting particles in a 2D potential.
In such systems, dynamic effects like phase locking phenomena \cite{Martinoli75, VanLook99, Kokubo02} and ratchet effects \cite{Villegas03a, Silva06} have been intensely investigated, as well as static effects like the formation of vortex quasicrystals \cite{Misko05, Misko06, Villegas06, Kemmler06, Kramer09} in particular for artificial defects arranged in a Penrose pattern.
An experimentally easily accessible quantity to gather integral information on the static interaction between a vortex lattice and a pinning array is the magnetic field dependent critical depinning current $I_c(B)$, where \textit{e.g.} commensurabilities between pinning sites and vortex lattices can be seen as pronounced maxima \cite{Fiory78, Moshchalkov98, Otani93, Martin97, Misko05, Misko06, Villegas06, Kemmler06, Kramer09}.
The implementation of artificial defects as pinning sites for Abrikosov vortices has furthermore relevance for a variety of applications, as in many cases the dissipative vortex motion leads to a reduction or a limitation of the performance of superconducting devices and hence is desired to be suppressed by pinning sites.
Approaches for improving the performance of superconducting microelectronic devices have found strategically positioned antidots feasible for the reduction of low-frequency flux noise in quantum interference devices \cite{Selders00} and for the reduction of vortex associated losses in coplanar microwave resonators \cite{Bothner11, Bothner12}.
Quasiperiodic arrays might be particularly suitable for some of these applications, as they have many built-in periodicities, which moreover are not only found for integer or rational but also for irrational multiples of the matching field, \textit{i.e.} the field for which the vortex density $n_v$ equals the density of pinning sites $n_p$.
For the Penrose tiling for instance, this leads to a significant broadening of the $I_c(B)$ peaks, as compared to the triangular defect lattice \cite{Misko05, Misko06, Kemmler06} and hence to a more homogeneous $I_c(B)$ dependence.
In this work, we experimentally investigate the $I_c(B)-T_c(B)$ phase boundary of superconducting niobium thin films with three different arrangements of antidots -- two kinds of quasiperiodic arrays, namely the Shield Tiling (ST) \cite{Gaehler88} and the Tuebingen Triangle Tiling (TTT) \cite{Baake90} -- and for reference a triangular lattice.
We perform our measurements close to $T_c$, where Little-Parks oscillations and collective Abrikosov vortex pinning are not strictly separable effects.
There might also be a smooth transition with decreasing $T$ from the wire network limit to a film with holes \cite{Patel07, Bothner12a}.
However, we are not going to focus on such a possible transition in this paper, but rather investigate commensurable states between an ensemble of quantized fluxoids and a given microtopology.
Hence, we monitor the phase boundary holistically.
We find and discuss signatures for commensurabilities at nonmatching excess flux densities, which have been predicted by theory \cite{Misko10} and which might be attributed to a caging effect and the formation of giant vortices.
The paper is organized as follows.
After the introductory Sec.~\ref{sec:Introduction}, we introduce the sample fabrication in Sec.~\ref{sec:Fabrication} and the method for their characterization in Sec~\ref{sec:Characterization}.
In Sec.~\ref{sec:Results} we present and discuss the experimental results on the phase boundary of niobium thin films with antidot arrays arranged in a Shield Tiling and a Tuebingen Triangle Tiling, respectively.
We also discuss experimental results on a sample with a hexagonal antidot array for reference.
Section~\ref{sec:Conclusion} contains the conclusion of this work.

\section{Sample Fabrication}

\label{sec:Fabrication}
\begin{figure}[h]
\centering \includegraphics{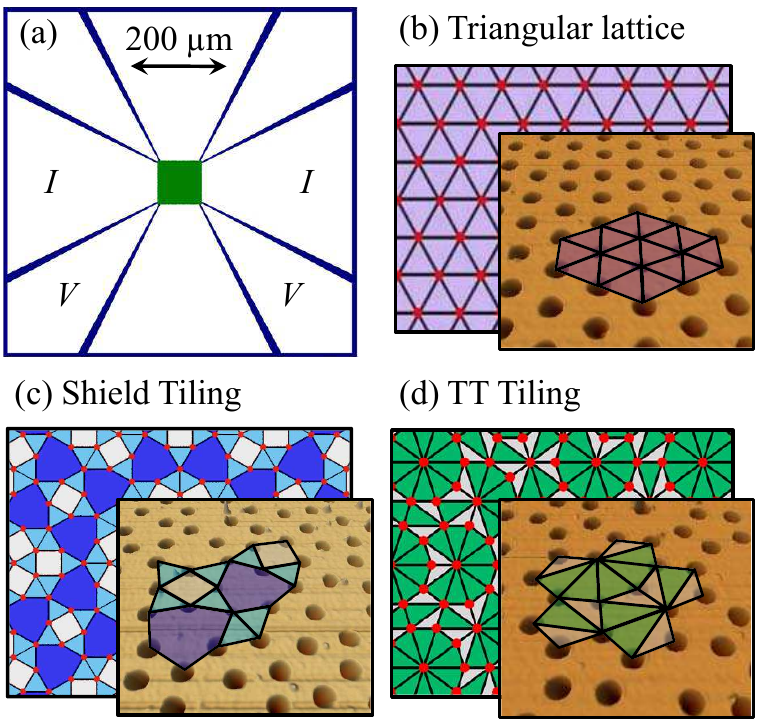}
\caption{(Color online) (a) Layout of a $800\times800$\,$\mu$m$^2$ chip with a cross shaped Nb structure for a four-probe current($I$)-voltage($V$) characterization of a $100\times100$\,$\mu$m$^2$ large area patterned with antidots (center square). (b)-(d) Sketches of the tilings and atomic force microscopy (AFM) images of the antidot arrays showing (b) the triangular/hexagonal lattice, (c) the Shield Tiling and (d) the Tuebingen Triangle Tiling; the antidot density of all three arrays is $n_p \approx 2\,\mu$m$^{-2}$ and the diameter of the circular antidots is $D\approx300\,$nm.}
\label{fig:Figure1}
\end{figure}

The experiments were carried out on cross-shaped structures patterned in $d=60\,$nm thick dc magnetron sputtered Nb films.
Figure~\ref{fig:Figure1} (a) shows a sketch of a $800\times800\,\mu$m$^2$ sample (white areas are Nb) with a $100\times100\,\mu$m$^2$ large center area (bridge), which is patterned with different arrangements of circular antidots.
Independent of the specific arrangement and array symmetry, the antidot density in the center area of these bridges is $n_p\approx2\,\mu$m$^{-2}$ corresponding to a total number of antidots $N_p\approx2\cdot10^4$ and matching field $B_1=n_p\Phi_0\approx4\,$mT.
The patterning of the structures including the antidots was performed by e-beam lithography and subsequent SF$_6$ reactive ion etching.
\begin{figure*}
\centering \includegraphics{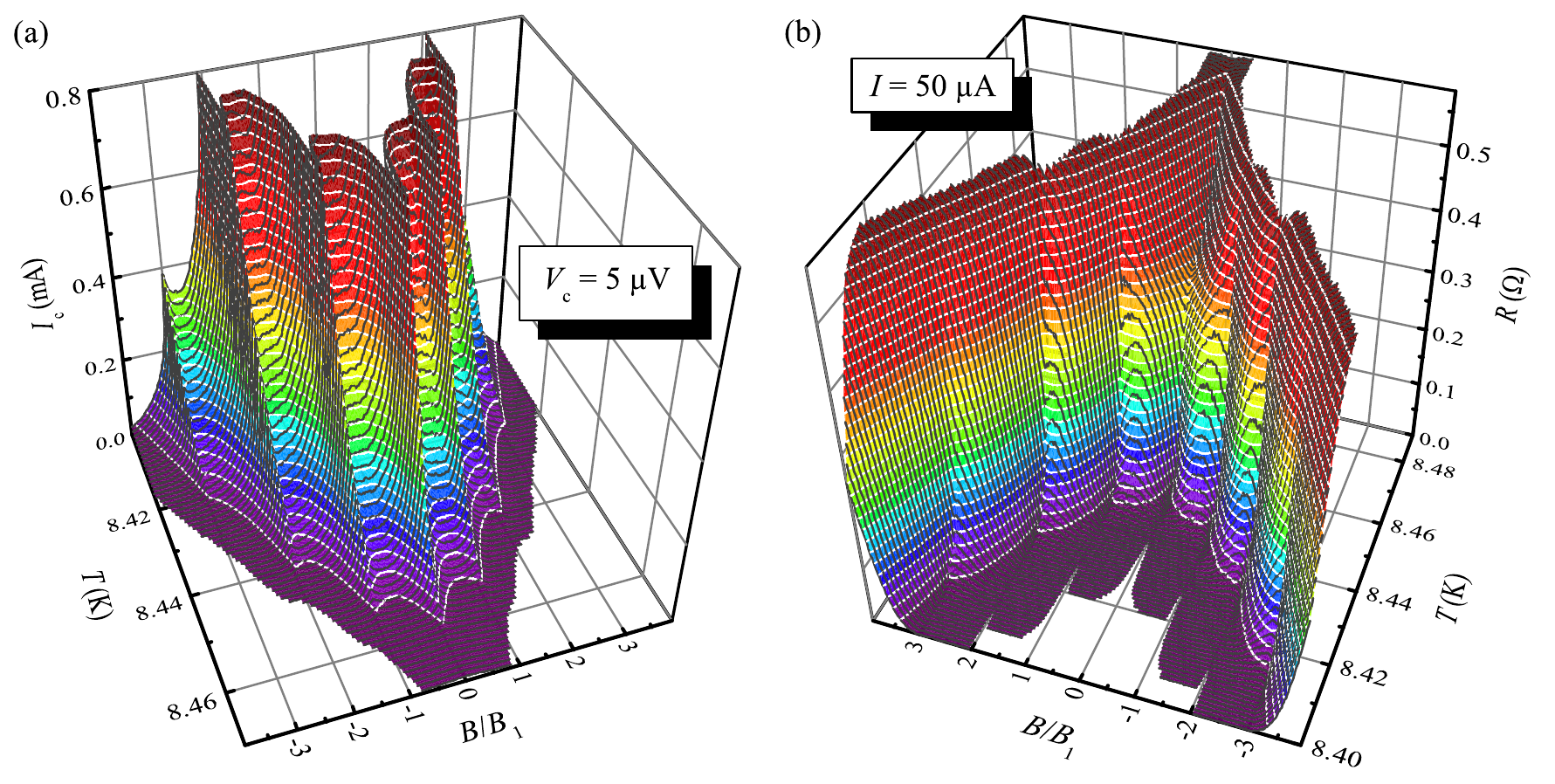}
\caption{(Color online) Data for Nb structure with \textbf{triangular} antidot lattice at variable temperature $T$ and magnetic flux density $B$ (normalized to the matching field $B_1\approx4\,$mT). (a) Critical current $I_c$ for a voltage criterion $V_c = 5\,\mu$V; (b) Resistance $R$ measured with a bias current $I=50\,\mu$A.}
\label{fig:Figure2}
\end{figure*}
Each bridge contains either a triangular/hexagonal array of antidots or quasiperiodically arranged antidots at the vertices of the Shield Tiling or the Tuebingen Triangle Tiling.
The three different patterns are shown in combination with a sketch of the tiling and an atomic force microscopy (AFM) image of the corresponding antidot array in Figs.~\ref{fig:Figure1} (b), (c) and (d).
A more detailed description and discussion of the quasiperiodic tilings is given together with the results in Sec.~\ref{sec:Results}.
The antidots of the samples shown here and used for the presented experimental results have a diameter $D\approx300\,$nm.
For a rough estimate of the magnetic penetration depth $\lambda$ and the coherence length $\xi$ in our samples, we determined the resistance slightly above $T_c$ as $R_n\approx 0.6\,\Omega$ (cf. Sec.~\ref{sec:Results}) and calculated the resistivity $\rho=R_nA/L=3\,\mu\Omega$cm with the center area cross section $A=100\,\mu$m$\,\times\,60\,$nm and length $L=100\,\mu$m.
From simulations we know that due to our cross-shaped geometry this resistivity has to be corrected by about a factor of two, as the bias current $I$ is not confined exactly to the center area.
So for further calculations we use the corrected resistivity $\tilde{\rho}=2\cdot\rho=6\,\mu\Omega$cm.
Using $\tilde{\rho} l=3.72\times10^{-6}\,\mu\Omega\,$cm$^2$ \cite{Mayadas72} we determine the electron mean free path $l\approx6\,$nm.
As we have neglected the antidots in this calculation, the value for $\tilde{\rho}$ is probably somewhat over-estimated and as a consequence the value for $l$ is somewhat under-estimated.
With the BCS coherence length for niobium of $\xi_0=39\,$nm and the ''dirty limit`` expression \cite{Tinkham} $\xi(T)=0.855\sqrt{l\xi_0}/\sqrt{1-T/T_c}$ we find $\xi(T)=13\,$nm$/\sqrt{1-T/T_c}$.
Analogously, with the London penetration depth $\lambda_L=38\,$nm at $T=0$ and $\lambda(T)=\sqrt{\xi_0/1.33l}\lambda_L/\sqrt{2(1-T/T_c)}$ we find $\lambda(T)=59\,$nm$/\sqrt{1-T/T_c}$.
As for all our data $T/T_c>0.98$ and hence $\lambda(T)>300\,$nm$\,\gg d=60\,$nm, we have to consider the thin film penetration depth $\Lambda=2\lambda^2/d$.
Again, $\xi(T)$ and $\lambda(T)$ are somewhat under- and over-estimated, respectively.

\section{Sample Characterization}

\label{sec:Characterization}

The samples were mounted in a low-temperature setup with high temperature variability $4.2\,$K$<T<100\,$K and stability $\delta T<1\,$mK \cite{AbsTemp}.
We record current-voltage characteristics (IVCs) over the whole 2D phase space of perpendicular magnetic field between the critical fields $\pm B_{\textrm{c2}}$ and temperatures close to the transition temperature $T_c$.
We keep the stepwidth between different temperatures $\Delta T$ and different values of magnetic field $\Delta B$ constant; so for each value of $T$ we have IVCs for all values of $B$ and vice versa.
This procedure enables us to extract all information as $I_c(B)$, $T_c(B)$ or $R(B)$-curves for arbitrary and freely selectable voltage criteria $V_c$, resistance criteria $R_c$ and bias currents $I$.
Figure~\ref{fig:Figure2} shows for a sample with a triangular antidot array (a) the critical current $I_c(T, B)$ and (b) the resistance $R(T, B)$.
From Fig.~\ref{fig:Figure2} (a), we can extract single vertical and horizontal slices, which correspond to $I_c(B)$-curves (dark lines) for constant $T$ and to $T_c(B)$-curves (bright lines) for constant $I$, respectively.
Similarly, from Fig.~\ref{fig:Figure2} (b) we get $R(B)$-curves for constant $T$ (dark lines) and $T_c(B)$-curves for constant $V$ (bright lines) by taking vertical and horizontal slices, respectively.
The cross shape of our Nb structures is expected to induce an inhomogeneous current density distribution across the center area.
This point has been particularly discussed for a ratchet experiment \cite{Gonzalez07, Silhanek08, Gonzalez08}, where the direction of the local driving force relative to the intrinsic symmetry axes of the pinning lattice is of great importance.
Also in periodic and quasiperiodic potential landscapes with circular antidots, the vortex dynamics might depend on the direction of the local transport current density \cite{Reichhardt99, Silhanek03, Villegas03, Reichhardt11}.
However, we do not find our experimental results to differ significantly for the two possible perpendicular directions of the bias current.
It is most likely anyway that the depinning current density is first reached directly between the voltage pad contact points, where the mean current density reaches its maximum due to the smallest cross section area.
So we think that the essential results of this work are not influenced by the bridge geometry.
Due to the used voltage amplifier, we had to face a voltage noise of several $10^{-7}\,$V.
To improve the signal to noise ratio and the visibility of small matching features, the results presented in this paper are not single-shot raw data but have been averaged and smoothed for their final interpretation and presentation.
First, we measured five IVCs at each $B-T$ phase space point, each consisting of  $1000$ data points, and averaged them for the further processing.
The averaged IVCs were slightly smoothed by a non weighted five-point adjacent-averaging before we extracted voltages for pre-defined currents or vice versa.
At the end we once more performed a non weighted three-point adjacent-averaging on the resulting $I_c(B)$, $T_c(B)$ and $R(B)$ slices.

\section{Experimental Results}

\label{sec:Results}
\subsection{The Triangular Lattice}

\begin{figure*}
\centering \includegraphics{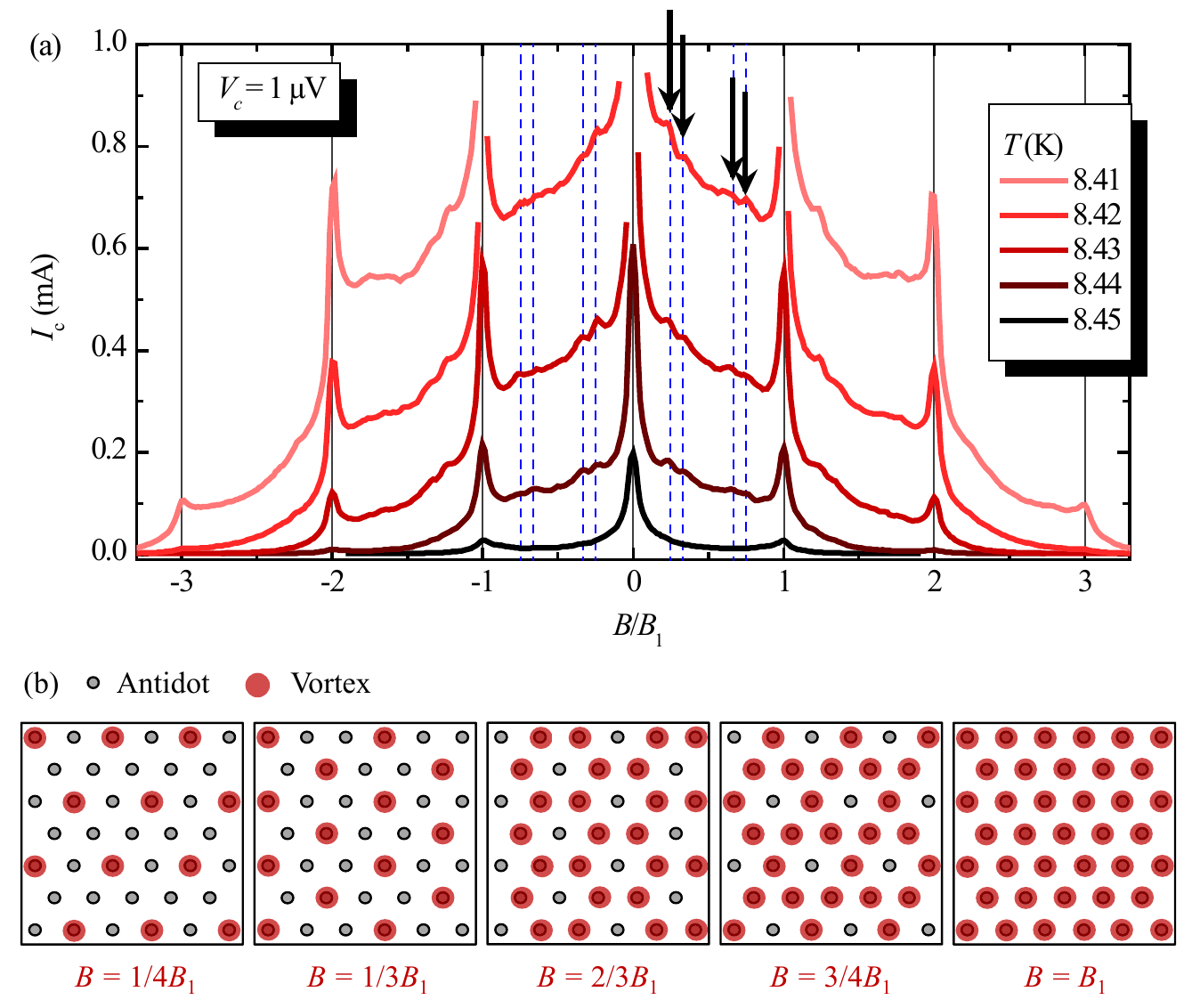}
\caption{(Color online) (a) Critical current $I_c$ vs normalized magnetic flux density $B/B_1$ of a Nb structure with a \textbf{triangular} antidot lattice at five different temperatures $T$; dashed vertical lines indicate the fractional flux densities $B/B_1=1/4, 1/3, 2/3$ and $3/4$ and arrows point to the corresponding local maxima. (b) Sketches of (possible) vortex configurations corresponding to local maxima of $I_c$ at $B/B_1=1/4, 1/3, 2/3, 3/4$ and $1$.}
\label{fig:Figure3}
\end{figure*}
Before we discuss our results on quasiperiodic antidot arrays, we present the results on the sample with the triangular lattice.
The data in Fig.~\ref{fig:Figure2} already reveal clear and strong features, which indicate commensurabilities between the flux line and antidot density.
We find pronounced and narrow ridges in (a) and canyons in (b) at $B=B_1$ (corresponding to one vortex per antidot).
We find quite similar but somewhat weaker structures for flux densities corresponding to two and three vortices per antidot ($B/B_1=2$ and $B/B_1=3$).
Our findings show that our samples and our data analysis are suitable to reliably detect not only strong commensurability features at integer matching fields, but also weaker ones at fractional matching flux densities.
These weaker commensurabilities at fractionals of the matching flux reveal themselves by characteristic fine-modulations of the phase boundary at flux densities between $B=0$ and $B_1$.
Figure~\ref{fig:Figure3} (a) shows several $I_c(B)$-slices.
Marked with arrows between $B=0$ and $B_1$, there are four small sub-peaks visible, which we attribute to the fractional filling factors $1/4$, $1/3$, $2/3$ and $3/4$ (dashed lines).
Corresponding vortex configurations and in addition the one for $B=B_1$ are depicted schematically in Fig.~\ref{fig:Figure3} (b).
The observed fractional matching peaks for $0<B<B_1$ and the corresponding vortex configurations have been found in theoretical as well as some experimental studies on triangular pinning lattices before, cf. \textit{e.g.} \cite{Reichhardt01, Ooi09}.
We find very similar peaks for flux densities between $B=B_1$ and $B=2B_1$,\textit{i.e.} for $B/B_1=5/4$, $4/3$, $5/3$ and $7/4$, indicating that the same configurations are stable for two-vortex occupations, that is when all antidots are already occupied by one vortex.
\begin{figure*}
\centering \includegraphics{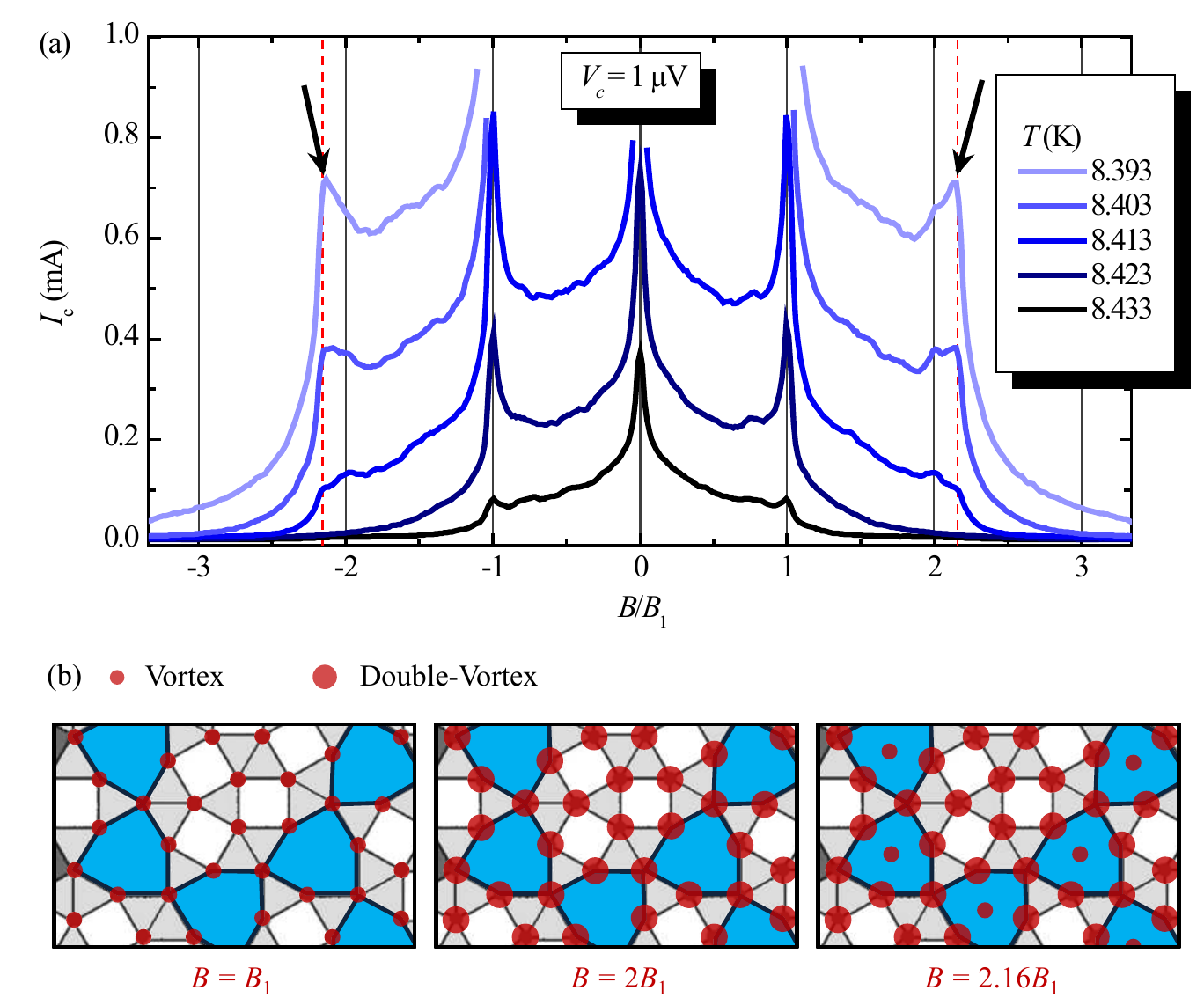}
\caption{(Color online) (a) Critical current $I_c$ vs normalized magnetic flux density $B/B_1$ of a Nb structure with a \textbf{Shield Tiling} lattice of antidots at five different temperatures $T$; dashed vertical lines indicate the fractional flux density $B/B_1=2.16$ and arrows point to the corresponding local maxima. (b) Sketches of (possible) vortex configurations corresponding to maxima of the critical current at $B/B_1=1, 2$ and $2.16$.}
\label{fig:Figure4}
\end{figure*}

\subsection{The Shield Tiling}
The Shield Tiling [cf. Fig.~\ref{fig:Figure1}~(c)] consists of three different tiles: a square, an equilateral triangle and a deformed hexagon with alternating interior angles of $\pi/2$ and $5\pi/6$.
This is a very interesting collection of tiles, as two of them locally resemble the energetically most favorable configurations of a periodic Abrikosov vortex lattice: the triangular and square array.
Each hexagon as the third basic shape leaves a rather large antidot-free area, what -- regarding the whole lattice -- leads to quasiperiodically distributed, individual cages or reservoirs for interstitial vortices as also described in Ref.~\cite{Misko10}.
The pattern was created by starting with six hexagonally arranged triangles followed by an iterative application of inflation rules \cite{Gaehler88}.
In a previous theoretical study on different quasiperiodic pinning arrays, we have found that the critical current density of a superconductor with antidots arranged in the Shield Tiling shows two pronounced $I_c(B)$ maxima.
One of them corresponded to the first matching field $B=B_1$.
The second maximum was found for an excess flux density of $B\approx1.18B_1$.
According to simulations, this corresponds to the situation, where all antidots are occupied by one vortex, and within each hexagon there is an additional interstitial vortex \cite{Misko10}.
These interstitial vortices are caged, \textit{i.e.} pinned by the repulsive interaction with the surrounding vortices, which are pinned at antidots.
In the sample investigated here, we expect the additional maximum at $B\approx1.16B_1$, which is at a slightly lower value than in Ref.~\cite{Misko10}.
This difference is due to the fact that the maximum is expected at $B=(1+N_h/N_p)B_1$, where $N_h$ and $N_p$ is the total number of hexagons and antidots, respectively.
The ratio $N_h/N_p$ converges for large numbers of antidots.
In the numerical approach in Ref.~\cite{Misko10} we have only used $N_p\approx500$ pinning sites with $N_h/N_p\approx0.18$.
In our experimentally used array with $N_p\approx20000$ antidots we have $N_h/N_p\approx0.16$.
Fig.~\ref{fig:Figure4} (a) shows $I_c(B)$-curves of the Shield Tiling sample at several different temperatures.
At first sight, there seems to be no additional maximum around $B=1.16B_1$ but only a very strong and single maximum of the critical current at $B=B_1$.
However, when we cool the sample to lower temperatures a double-peak structure appears with one peak at $B=2B_1$ and one at a slightly higher field value $B\approx2.16B_1$.
This additional peak -- or more precisely speaking its height relative to the one at $B=2B_1$ -- has a strong dependence on temperature.
It is almost absent at $T=8.413\,$K but then grows and becomes the dominating peak at $T=8.393\,$K (and probably below).
A very similar relation and development between two peaks with temperature has been reported for caged vortices in periodic pinning arrays before \cite{Berdiyorov06} and might be attributed to an increasing interaction strength between the vortices due to a decreasing penetration depth.
The predicted matching at an excess vortex density peak seems to appear not at $B=1.16B_1$ but at $B=2.16B_1$.
It is thus very likely that singly occupied antidots are still too attractive for vortices entering above $B=B_1$, hence, they first occupy the antidots twice before the centers of the hexagons become energetically more favorable.
We show sketches of the proposed vortex configurations in Fig.~\ref{fig:Figure4} (b).
Of course, the situation depicted here is somewhat idealized.
In simulations we find similar situations, but we also find a certain degree of disorder in the vortex lattice in any case.
For example, for $B=2B_1$ it might well be that not all six antidots of the hexagons are doubly occupied.
It is rather likely that one or two of them only host one vortex, while there are also interstitials present.
The simulations we refer to here, are not shown, as they were not performed exactly with the experimental parameters.
However, they served as an inspiration for the proposed vortex configurations.
The discussed observations of commensurabilities in the Shield tiling are closely related to widely discussed caging effects in different kinds of periodic pinning arrays, including randomly or periodically diluted triangular lattices \cite{Berdiyorov06, Reichhardt07, Kemmler09, Cao11, Latimer12}.
These studies have revealed, that matching peaks can be shifted to higher field values than those, for which vortex and antidot density are equal.
Whenever there are areas with missing antidots, these areas fill up with interstitials, when the antidots are saturated and these configurations with interstitials can be more stable than the configurations without interstitials.
\begin{figure*}
\centering \includegraphics{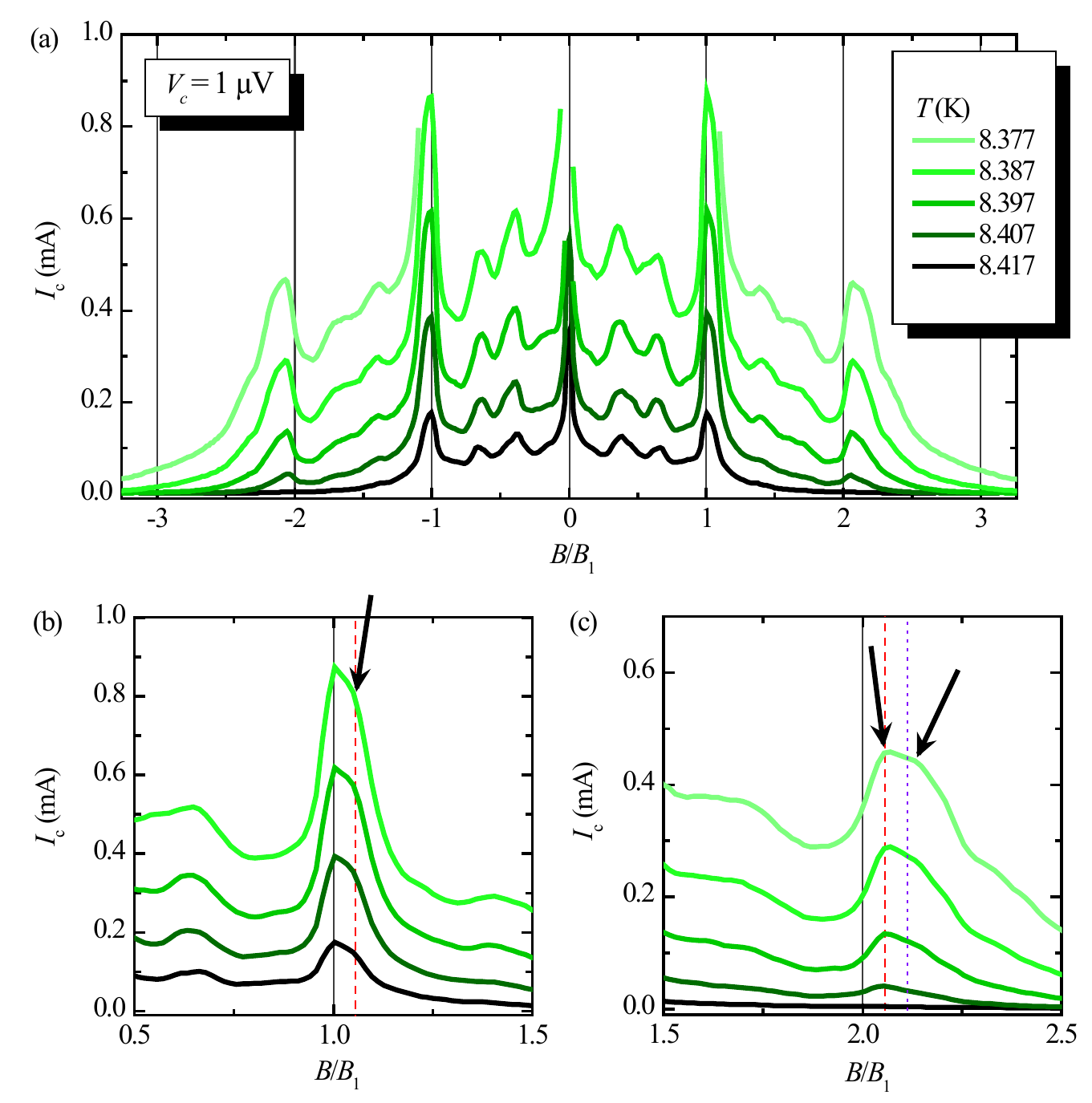}
\caption{(Color online) (a) Critical current $I_c$ vs normalized magnetic flux density $B/B_1$ of a Nb structure with a \textbf{Tuebingen Triangle Tiling} lattice of antidots at five different temperatures $T$; (b) and (c) show  $I_c(B)$ in more detail at $B\approx B_1$ and $2B_1$, respectively. Arrows point to the excess flux commensurabilities; dashed and dotted vertical lines mark the flux densities $B/B_1=1+1/\tau ^6$ ((b), dashed), $B/B_1=2+1/\tau ^6$ ((c), dashed) and $B/B_1=2+2/\tau ^6$ ((c), dotted).}
\label{fig:Figure5}
\end{figure*}

\subsection{The Tuebingen Triangle Tiling}

The second quasiperiodic arrangement -- the Tuebingen Triangle Tiling -- is depicted in Fig.~\ref{fig:Figure1} (d).
It consists of two different isosceles triangles, one acute-angled and one obtuse angled.
This tiling can also be constructed by an iterative application of inflation rules \cite{Baake90} after a starting configuration of ten acute-angled triangles.
In contrast to the triangular and the Shield tiling, the edge lengths are not equal for all basic shapes here, but we find two different lengths, whose ratio is the golden mean $\tau\approx1.618$.
This can lead to a locally reduced pinning site density, when ten triangles are composed such that 10 antidots form a ring, for example in the left upper and lower corner of Fig.~\ref{fig:Figure1} (d).
Below, these full rings with 10 antidots surrounding one as a circle are called 10/10-circles.
The 10/10-circles have a high antidot line density on the circle and a reduced density in the interior and may again be viewed as cages or reservoirs for interstitial vortices.
In contrast to the cages in the Shield Tiling we here find the situation that there actually is one antidot in the center of each 10/10-circle.
Analogously to the Shield Tiling discussion we have to calculate the ratio of the number of 10/10-circles $N_\textrm{10}$ over the total number of antidots $N_p$ in order to assign the appearance of local maxima in $I_c(B)$ to vortex configurations related to the circles.
This ratio is found to be $N_\textrm{10}/N_p=1/\tau^6\approx0.056$ for $N_p \to \infty$, as after each iteration the total number of antidots increases by a factor of $\tau^2$, \textit{i.e.} $N^{i}_{p}=\tau^2N^{i-1}_p$ for the $i$th iteration.
After three iterations each antidot has transformed into a 10/10-circle, \textit{i.e.} $N^i_\textrm{10}=N^{i-3}_p=(1/\tau^6)N^i_p$.
The next step is to investigate the phase boundary of the TTT sample with respect to modulations associated with $\tau^6$.
\begin{figure}
\centering \includegraphics{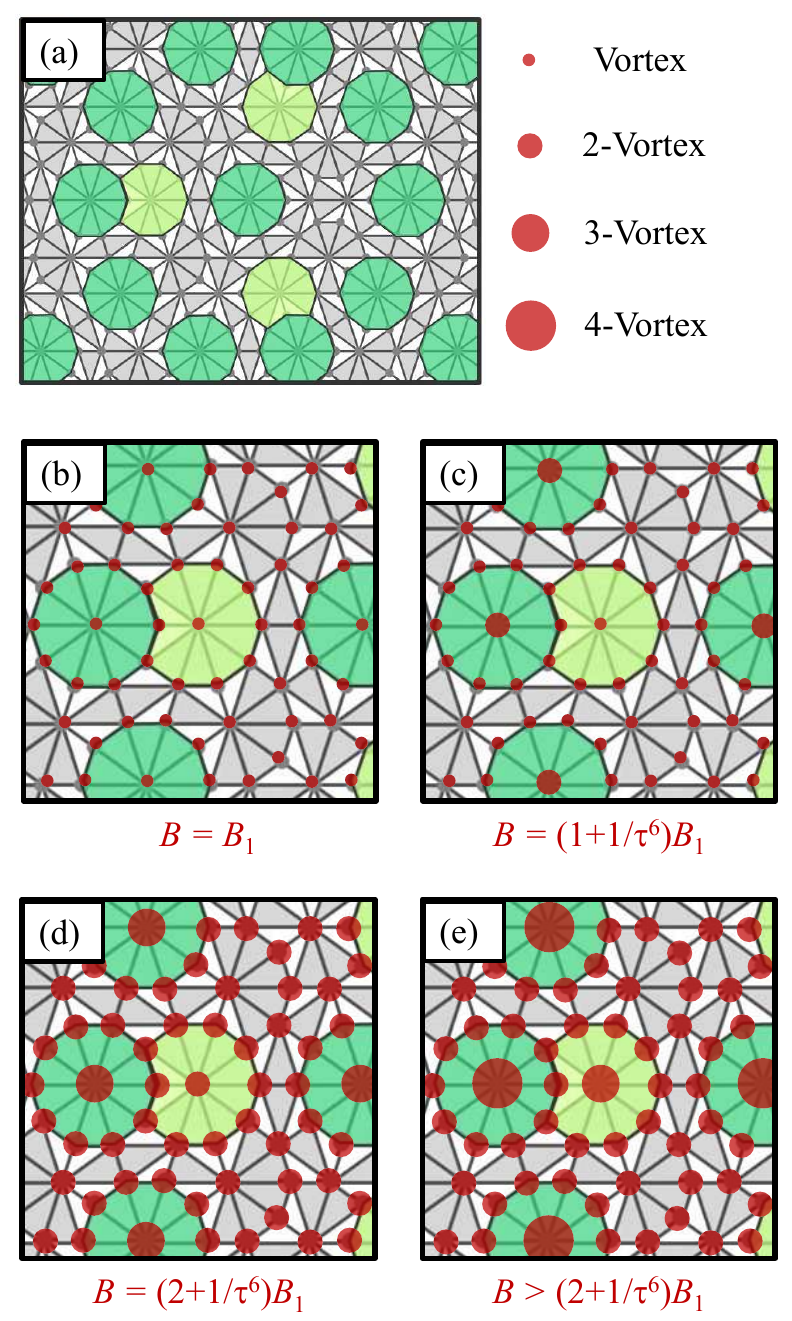}
\caption{(Color online) (a) Sketch of a part of the Tuebingen Triangle Tiling with colored 10/10-circles and 9/10-circles. (b)-(e) Possible vortex configurations related to the critical current maxima at (b) $B=B_1$, (c) $B=(1+\tau^6)B_1$, (d) $B=(2+\tau^6)B_1$ and (e) $B>(2+\tau^6)B_1$ including higher order giant vortices.}
\label{fig:Figure6}
\end{figure}
Figure~\ref{fig:Figure5} (a) shows $I_c(B)$-curves of the Tuebingen Triangle Tiling sample at five different temperatures.
There are two strong sub-peaks for flux densities $B\approx0.39B_1$ and $B\approx0.65B_1$.
Similarly to the triangular lattice this substructure for (probably irrational) fractionals of $B_1$ repeats itself with a somewhat smaller amplitude between $B=B_1$ and $2B_1$.
Also in our previous numerical simulations we have found this characteristic double-peak structure at approximately the same values as in the present experimental study \cite{Misko10}.
However, we have not found a simple connection between these peaks and intuitively understandable vortex configurations.
Regarding the first matching field, there seems to be just the standard single maximum for $B=B_1$ at first sight.
But a zoom-in as in Fig.~\ref{fig:Figure5} (b) indicates a double peak structure, where the two peaks are merged into a broad one with two kinks at $B=B_1$ and at a slightly higher value (marked by arrow).
The vertical dashed line in Fig.~\ref{fig:Figure5} (b) indicates the position of $B=(1+1/\tau^6)B_1$ and matches very well the flux density value of the second kink.
The second peak hence can be attributed to the flux value, where we have one vortex per antidot and one additional vortex per 10/10-circle in the sample.
One possibility is that the additional vortex is at an interstitial position somewhere inside the circle or that the additional vortex pushes the vortex out of the center antidot, so that the two of them symmetrically occupy the circle interior as interstitials.
However, we favour a third possibility.
The idea is connected to a recent imaging experiment of vortices in a quasiperiodic Penrose pinning lattice \cite{Kramer09}, which revealed a symmetry-induced formation of giant vortices.
In the Tuebingen Triangle Tiling situation we think that the antidot in the center of the 10/10-circles is occupied by a double-vortex at $B=(1+1/\tau^6)B_1$, while all other pinning sites are occupied by one vortex.
This double-vortex is supposed to compensate the magnetic pressure from the surrounding vortices on the ten vortices pinned at the circle line.
It stabilizes the whole configuration this way.
Exactly this vortex configuration with ten vortices forming a circle and one double-vortex in its center was found in Ref. \cite{Kramer09} for pinning sites arranged in the Penrose tiling.
It is remarkable that in the Penrose pattern the pinning site circle is not even complete, but has three pinning site free positions.
The fact that the giant vortex configuration forms anyway, strongly indicates its energetic favorability.
This should be even more favorable in the Tuebingen Triangle Tiling.
Figure~\ref{fig:Figure6} (a) depicts the Tuebingen Triangle Tiling and possible vortex configurations corresponding to $B=B_1$ (Fig.~\ref{fig:Figure6} (b)) and $B=(1+1/\tau^6)B_1$ (Fig.~\ref{fig:Figure6} (c)).
Similarly to the phase boundary of the Shield Tiling, we also find interesting features in $I_c(B)$ of the Tuebingen Triangle Tiling sample around $B=2B_1$, cf. Fig.~\ref{fig:Figure5} (c).
We note the surprising absence of a maximum at $B=2B_1$.
Nevertheless we find a clear double peak structure similar to the one around the first matching field.
The first of the two peaks is approximately at a value $B=(2+1/\tau^6)B_1$ (dashed vertical line) and the second at a somewhat higher value, but not exactly at $B=(2+2/\tau^6)B_1$ (dotted line).
The interpretation of the first peak in terms of vortex configurations is obvious.
Here we have the situation that all antidots are occupied by two vortices, whereas those antidots in the center of a 10/10-circle are occupied by three vortices, cf. Fig.~\ref{fig:Figure6} (d).
The absence of the $2B_1$ peak indicates that a third vortex in the center antidot might be necessary to stabilize the configuration.
In contrast to the situation around $B=B_1$, it is likely that some vortices from the circle lines sit in the interior as interstitials rather than being pinned at an antidot.
We think this situation is similar to that in the Shield tiling, where the caging peak gets stronger than the $2B_1$ peak with decreasing temperature.
Analogous to the peak at $B=(1+1/\tau^6)B_1$ we should find a maximum for the situation when the antidots in the 10/10-circle center are occupied by twice the number of vortices as the surrounding pinning sites, \textit{i.e.} by a quadruple-vortex.
But this seems not to be the case, as the second kink of the double peak is found at a higher value than $B=(2+2/\tau^6)B_1$ (dotted vertical line).
We believe, that in this second peak another structure in the Tuebingen Triangle Tiling comes into play, which we call the 9/10-circle.
This structure is almost identical to the 10/10-circle but with one of the ten circle antidots sitting at a closer position to the one in the center.
In Fig.~\ref{fig:Figure6} (a) three 9/10-circles are visible.
They are always directly neighboring a 10/10-circle forming an 8-like structure.
One possibility to get a stable vortex configuration is sketched in Fig.~\ref{fig:Figure6} (e), where the center antidots of the 10/10-circles are occupied by quadruple-vortices, the center antidots of the 9/10-circles are occupied by triple-vortices and all other by a double-vortex.
Unfortunately we cannot give an analytical expression for the corresponding flux density value, as we were not able to determine the ratio of the number of 9/10-circles over the total number of antidots.
Probably one has to consider more complicated configurations including interstitial vortices and also present 8/10-circles to find the true vortex configurations.
Furthermore, there seem to be some more small shoulders and peaks above $2B_1$, which we cannot unambiguously attribute to simple vortex configurations.

\subsection{The Transition Temperatures}

We finally consider selected horizontal phase boundary $T_c(B)$ slices of the three samples.
We define $T_c$ by a fixed ratio $R_c/R_n$, where $R_n$ is the normal state resistance slightly above the superconducting transition at $B=0$ and $R_c$ is a resistance value within the transition regime.
\begin{figure}
\centering \includegraphics{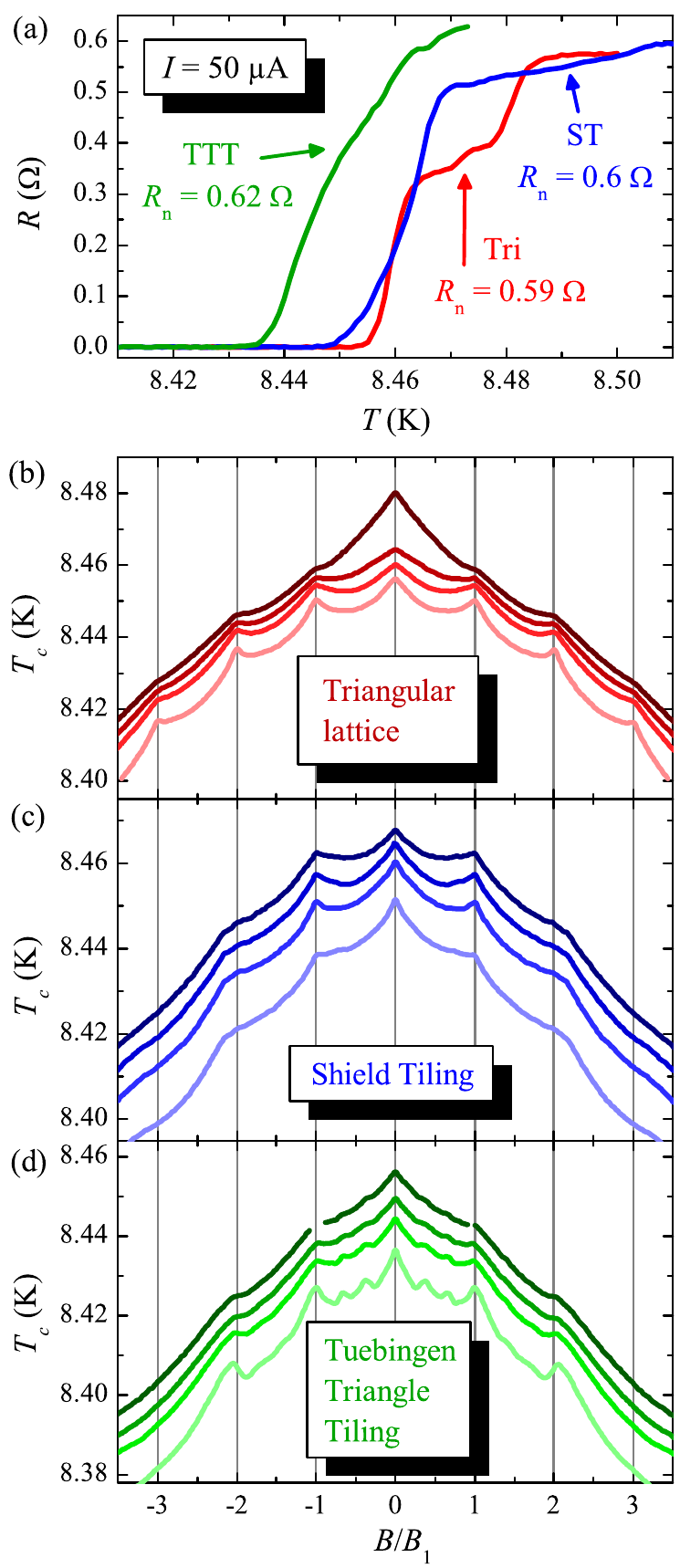}
\caption{(Color online) (a) Resistance $R$ vs temperature $T$ close to the transition temperature $T_c$ of the three investigated samples; (b)-(d) Transition temperature $T_c$ vs normalized magnetic flux density $B/B_1$ for four different resistance criteria (from top to bottom in each graph $R_c/R_n=0.7, 0.5, 0.3, 0.1$) of the three investigated samples; $R_n$ was determined at $T=8.6\,$K.}
\label{fig:Figure7}
\end{figure}
The transition temperatures of the three samples are comparable, cf. Fig.~\ref{fig:Figure7} (a).
However, the absolute values as well as their ratios depend on the resistance criterion.
For instance, the sample with the triangular antidot lattice shows a clear two-step (or even three-step) transition $R(T)$ for an unknown reason and depending on the criterion has a smaller \textit{or} larger critical temperature than the ST sample.
Also, the general shape of the transition (e.g. the curvature of the curve at $R_c/R_n=0.5$) slightly varies from sample to sample and all transitions show smaller or larger steps.
Fingerprints of these stepwise transitions can also be found in the $T_c(B)$ curves, which are shown in Fig.~\ref{fig:Figure7} (b)-(d).
Concerning our discussion of excess flux stabilities, the most important observation in the $T_c(B)$ data is that these commensurabilities for the Shield and the Tuebingen Triangle Tiling, which we have already found in the $I_c(B)$ patterns, are also clearly visible in $T_c(B)$.
All commensurability features are rather weak for the highest resistance criterion $R_c/R_n=0.7$, but get stronger and more pronounced with decreasing $R_c$.
One exception with yet unclear origin has been observed for the Shield Tiling, where the first matching peak seems to get weaker again for the lowest criterion $R_c/R_n=0.1$.
When $R_c$ is decreased further (and with it the temperature), this peak gets stronger again, cf. also $I_c(B)$ in Fig.~\ref{fig:Figure4}, where the temperatures are lower than in Fig.~\ref{fig:Figure7} (c).
This behaviour of the first ST matching peak strength might be connected to the transition from a superconducting wire network to a thin film with holes, but the observed variations and steps in the transition as well as the nature of the problem confront us with several complications for a detailed interpretation and quantitative analysis. 
Due to a not well-defined transition temperature, we can not reliably calculate the values for the characterictic length scales such as penetration depth $\Lambda(T)$ and coherence length $\xi(T)$, which diverge at $T_c$ and hence are strongly dependent on the measurement temperature as well as on the transition temperature.
It is thus not possible to reliably calculate, at which temperature the coherence length $\xi(T)$ is comparable to the width $w$ of the superconducting material in between the holes, \textit{i.e.} at which temperature the transition from a superconducting wire network ($\xi>w$) to a film with holes ($\xi<w$) takes place. 
For the case of a quasiperiodic arrangement of antidots, this transition is not even well-defined, as $w$ is spatially varying.
Nevertheless, for having an idea of the relevant numbers, we can calculate the approximate temperature below $T_c$, at which the distance between the holes is comparable to the coherence length.
Assuming $T_c=8.5\,$K and $w\approx500\,$nm we find with the expressions from Sec.~\ref{sec:Fabrication} that $\xi>500\,$nm if $T>8.494\,$K.
In other words, as long as we are farther away from $T_c$ than $6\,$mK the samples can be viewed as superconducting films with holes rather than as wire networks.
Although the numbers here are only rough estimates, all $I_c(B)$ curves above can be considered to have been taken at temperatures well below $T_c$ and so the observations correspond to Abrikosov vortex physics.
However, due to the differences in the shape of the $R(T)$ transition and the absolute values of $T_c$ we cannot compare the performance of different pinning lattices concerning critical currents, neither at absolute values of $T$ nor at reduced values $T/T_c$.

\section{Conclusions}

\label{sec:Conclusion}
In conclusion, we have experimentally investigated the $I_c(B)-T_c(B)$ phase boundary of niobium thin films with quasiperiodic microtopology.
We have done this by means of transport characterization measurements and discussed the results with the focus on observed commensurability effects for flux densities larger than the matching flux.
Our experiments confirm some theoretically predicted and yet experimentally unobserved excess flux matching effects for antidots arranged corresponding to the Shield Tiling and to the Tuebingen Triangle Tiling.
For the Shield Tiling, we attribute this excess flux matching to an effective caging of Abrikosov vortices in quasiperiodically distributed antidot-free hexagons.
We find similar excess flux commensurability features for antidots arranged in the Tuebingen Triangle Tiling.
However, apart from the possibility of caging, the results can be understood as a consequence of recently reported symmetry-induced formation of higher order giant vortices.
In particular, we find in both cases that the additional vortices, which create the excess flux commensurabilities, are needed to maximize the local stability of the vortex lattice.
To finally confirm our interpretations of the matching effects and to decide which vortex configurations are correct, one needs to carry out imaging experiments such as magneto-optical imaging, Bitter decoration or Hall probe microscopy.
This work has been supported by the European Research Council via SOCATHES and by the Deutsche Forschungsgemeinschaft via the SFB/TRR 21.
DB gratefully acknowledges support by the Evangelisches Studienwerk e.V. Villigst.
MK gratefully acknowledges support by the Carl-Zeiss Stiftung.
VRM gratefully acknowledges support by the ``Odysseus'' Program of the Flemish Government and the Flemish Science Foundation (FWO-VI).
The authors thank Franco Nori for fruitful discussions on quasiperiodic pinning arrays.


\begin{thebibliography}{26}
\bibitem{AbsTemp}
  The absolute values of $T$ have an uncertainty of $\pm0.5\,$K due to the uncertainty in the calibration of the temperature sensor.
\end{thebibliography}
\end{document}